\documentclass[aps,twocolumn,prd,10pt,showpacs,showkeys,preprintnumbers,superscriptaddress,floatfix,longbibliography,nofootinbib]{revtex4-1}

\pdfoutput=1
\usepackage{amsmath,amsfonts,amssymb,mathrsfs,graphicx,color,longtable}
\usepackage{hyperref}
\usepackage{graphicx}
\usepackage{amsfonts,amsmath,amssymb,bm}
\usepackage{array}
\usepackage{color}
\usepackage{enumitem}
\usepackage[explicit]{titlesec}
\usepackage[usenames,dvipsnames,table]{xcolor}
\usepackage[utf8]{inputenc}
\usepackage{comment}
\usepackage[letterpaper, portrait, margin=0.7in]{geometry}
\usepackage{booktabs} 
\allowdisplaybreaks

\newcommand\bout{\bgroup\markoverwith{\textcolor{blue}{\rule[0.5ex]{4pt}{0.8pt}}}\ULon}

\newcolumntype{C}[1]{>{\centering\let\newline\\\arraybackslash\hspace{4pt}}m{#1}}

\begin{document}

\title{Back to (Mass-)Square(d) One:\\ The Neutrino Mass Ordering in Light of Recent Data}

\author{Kevin J. Kelly}
\email{kkelly12@fnal.gov}
\thanks{\scriptsize \!\! \href{https://orcid.org/0000-0002-4892-2093}{0000-0002-4892-2093}}
\affiliation{Theoretical Physics Department, Fermi National Accelerator Laboratory, P. O. Box 500, Batavia, IL 60510, USA}

\author{Pedro A. N. Machado}
\email{pmachado@fnal.gov}
\thanks{\scriptsize \!\! \href{https://orcid.org/0000-0002-9118-7354}{0000-0002-9118-7354}}
\affiliation{Theoretical Physics Department, Fermi National Accelerator Laboratory, P. O. Box 500, Batavia, IL 60510, USA}

\author{Stephen J. Parke}
\email{parke@fnal.gov}
\thanks{\scriptsize \!\! \href{https://orcid.org/0000-0003-2028-6782}{0000-0003-2028-6782}}
\affiliation{Theoretical Physics Department, Fermi National Accelerator Laboratory, P. O. Box 500, Batavia, IL 60510, USA}

\author{Yuber F. Perez-Gonzalez}
\email{yfperezg@northwestern.edu}
\thanks{\scriptsize \!\! \href{https://orcid.org/0000-0002-2020-7223}{0000-0002-2020-7223}}
\affiliation{Theoretical Physics Department, Fermi National Accelerator Laboratory, P. O. Box 500, Batavia, IL 60510, USA}
\affiliation{Department of Physics \& Astronomy, Northwestern University, Evanston, IL 60208, USA}
\affiliation{Colegio de F{\'i}sica Fundamental e Interdisciplinaria de las Am{\'e}ricas (COFI), 254 Norzagaray Street, San Juan, Puerto Rico 00901}

\author{Renata Zukanovich Funchal}
\email{zukanov@if.usp.br}
\thanks{\scriptsize \!\! \href{https://orcid.org/0000-0001-6749-0022}{0000-0001-6749-0022}}
\affiliation{Instituto de F{\'i}sica, Universidade de S{\~a}o Paulo, C.P. 66.318, 05315-970 S{\~a}o Paulo, Brazil}

\date{\today}

\preprint{FERMILAB-PUB-20-330-T}

\begin{abstract}
We inspect recently updated neutrino oscillation data -- specifically coming from the Tokai to Kamioka and NuMI Off-axis $\nu_e$ Appearance experiments -- and how they are analyzed to determine whether the neutrino mass ordering is normal ($m_1 < m_2 < m_3$) or inverted ($m_3 < m_1 < m_2$). 
We show that, despite previous results giving a strong preference for the normal ordering, with the newest data from T2K and NOvA, this preference has all but vanished. 
Additionally, we highlight the importance of this result for non-oscillation probes of neutrinos, including neutrinoless double beta decay and cosmology. 
Future experiments, including JUNO, DUNE, and T2HK will provide valuable information and determine the mass ordering at a high confidence level.
\end{abstract}

\maketitle


\textbf{Introduction.} --- 
By observing the phenomenon of neutrino oscillations, we have determined a number of their properties. Current data allow us to understand how neutrinos mix and that there are two non-zero mass scales. Neutrinos in any oscillation environment are highly relativistic, so experiments are only sensitive to differences of masses squared, $\Delta m_{ji}^2 \equiv m_j^2 - m_i^2$, between the three neutrino mass eigenstates\footnote{Defined where $\nu_1$ ($\nu_3$) has the largest (smallest) admixture of $\nu_e$.} $\nu_i$, with masses $m_i$.

Solar and reactor neutrino experiments, have determined\footnote{Barring additional new physics in the neutrino sector~\cite{Coloma:2016gei,Coloma:2017egw,Coloma:2017ncl}.} $\Delta m_{21}^2 \approx +7.5 \times 10^{-5}~{\rm eV}^2$~\cite{Aharmim:2011vm, Gando:2013nba}. Accelerator/atmospheric neutrinos have determined $\left\lvert \Delta m_{31}^2\right\rvert \approx 2.5 \times 10^{-3}~{\rm eV}^2 \gg \Delta m_{21}^2$ but, in general, are not sensitive to the sign of $\Delta m_{31}^2$ -- this is the neutrino mass ordering (MO) problem -- whether nature prefers $m_1 < m_2 < m_3$, the normal ordering (NO), or $m_3 < m_1 < m_2$, the inverted ordering (IO)~\cite{deSalas:2018bym}.

There are two ways to determine the MO, utilizing interference or matter effects. The first, employed by the upcoming JUNO~\cite{An:2015jdp} experiment, relies on measuring neutrino oscillations where both mass-squared-splittings are relevant. Alternatively, accelerator neutrino experiments where effects due to $\Delta m_{31}^2$ are dominant and matter effects (from neutrino interactions with rock along their journey) are relevant, are also sensitive to the MO. 
A combination of measuring oscillation probabilities for muon-neutrino disappearance $P(\nu_\mu \to \nu_\mu)$ and electron-neutrino appearance $P(\nu_\mu \to \nu_e)$ (as well as the corresponding probabilities for antineutrinos) allows for long-baseline experiments to measure the MO. However, degeneracies exist between determining the MO, the atmospheric octant, and the degree of CP violation in the leptonic sector ($\delta_{\rm CP}$).

The latter strategy is employed by the currently-operating Tokai to Kamioka (T2K)~\cite{Abe:2018wpn,Abe:2019vii,Abe:2019ffx} and NuMI Off-axis $\nu_e$ Appearance (NOvA)~\cite{Adamson:2017gxd,NOvA:2018gge,Acero:2019ksn} experiments, which measure $P(\nu_\mu \to \nu_e)$ and $P(\overline{\nu}_\mu \to \overline{\nu}_e)$ at long distances. Super-Kamiokande (SK)~\cite{Abe:2017aap} also has modest sensitivity to the MO by studying atmospheric neutrino oscillations, where matter effects are important. 

\begin{figure*}
\begin{center}
\includegraphics[width=0.6\linewidth]{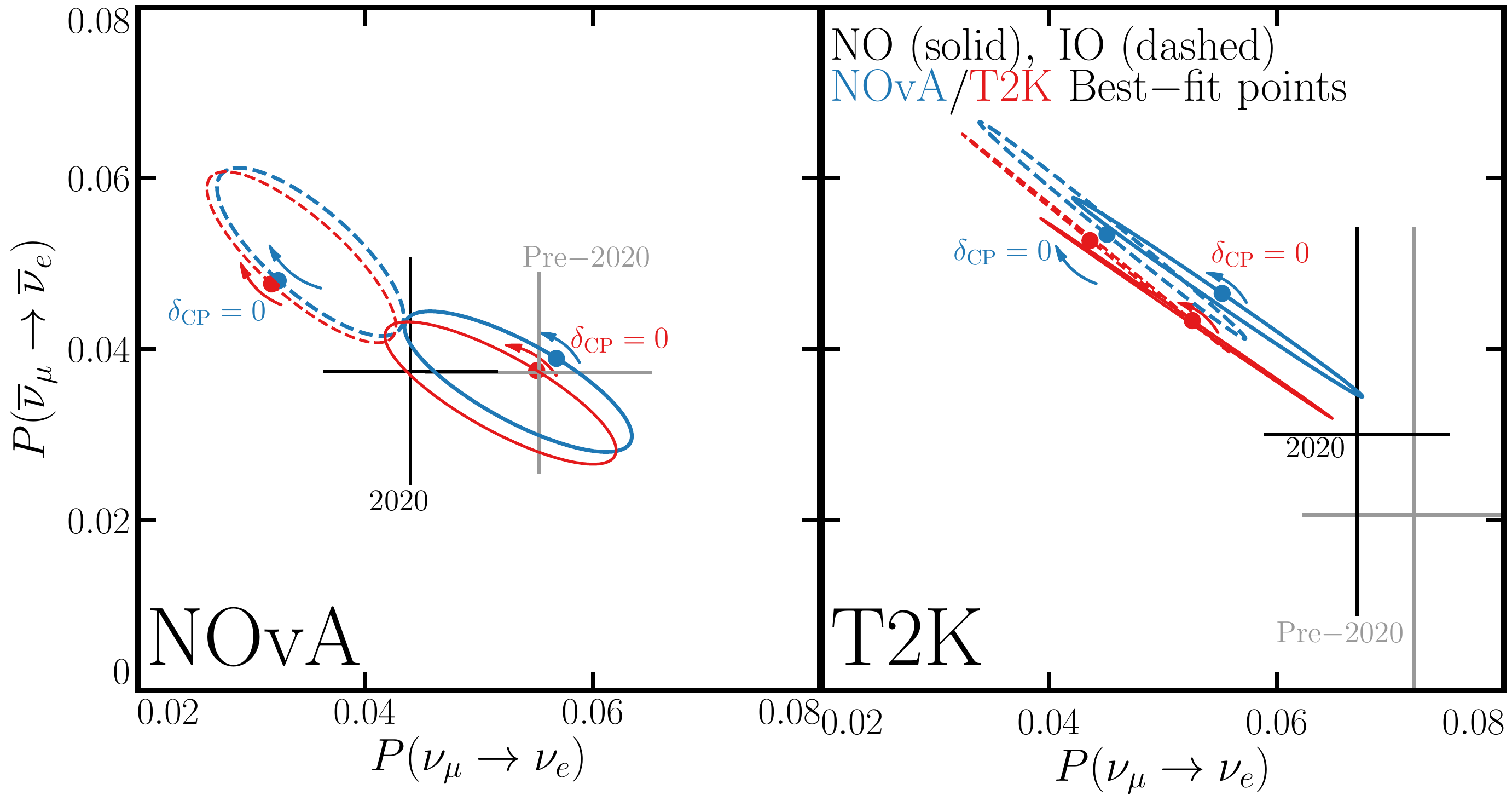}
\caption{Bi-probability plots of oscillation probabilites for neutrinos (x-axes) and antineutrinos (y-axes) at the $L$ and $E$ for NOvA (left panel) and T2K (right panel) while varying $\delta_{\rm CP}$. Black (grey) crosses indicate measurements with statistical uncertainty only for the two experiments using their 2020 (pre-2020) results.
 Ellipses correspond to best-fit points according to NOvA (blue) and T2K (red) fits under the the Normal (solid) or Inverted (dashed) mass ordering. 
Dots denote probabilities for $\delta_{\rm CP} = 0$, with arrows indicating increasing $\delta_{\rm CP}$.
\label{fig:BiProbPP}}
\end{center}
\end{figure*}
As of mid-2020, existing data, driven largely by these three experiments, exhibited a strong preference for the NO over the IO:  $\Delta\chi^2_{\rm (NO,IO)} \equiv \chi^2_{\rm min,IO} - \chi^2_{\rm min,NO} \approx 10$~\cite{Esteban:2018azc,Capozzi:2020qhw,deSalas:2020pgw}. However, T2K, NOvA, and SK have each provided preliminary updates~\cite{T2KNu2020,NOvANu2020,SKNu2020}. We will demonstrate that this NO preference vanished due to correlations between the data, as well as the degeneracies between MO, octant, and $\delta_{\rm CP}$. Additionally, we will discuss the ramifications of this result and provide some outlook for the future.


\textbf{MO sensitivity at Long-Baseline Experiments.} ---
In long-baseline experiments like T2K and NOvA (and the planned T2HK~\cite{Abe:2018uyc} and DUNE~\cite{Abi:2020evt,Abi:2020qib} experiments), oscillations due to $\Delta m_{21}^2$ have yet to develop, so the expansion parameter $\Delta m_{21}^2 L/4E$ is perturbatively small. Assuming neutrinos propagate through constant-density matter, the oscillation probability of $\nu_\mu \to \nu_e$ (with energy $E$ and distance $L$) can be approximated~\cite{Nunokawa:2007qh}
\begin{align}
&P_{\mu e} \equiv P(\nu_\mu \to \nu_e) \approx 4s_{23}^2 s_{13}^2 c_{13}^2 \frac{\sin^2\left(\Delta_{31} - aL\right)}{\left(\Delta_{31}-aL\right)^2}\Delta_{31}^2 \nonumber \\
&+8\frac{J}{\sin\delta_{\rm CP}} \frac{\sin\left(\Delta_{31}-aL\right)}{\left(\Delta_{31}-aL\right)} \Delta_{31} \frac{\sin{(aL)}}{(aL)}\Delta_{21} \cos{\left(\Delta_{31}+\delta_{\rm CP}\right)} \nonumber \\
&+4s_{12}^2 c_{12}^2 c_{13}^2 c_{23}^2 \frac{\sin^2\left(aL\right)}{\left(aL\right)^2}\Delta_{21}^2, \label{eq:PmeMatter}
\end{align}
where $\Delta_{j1} \equiv \Delta m_{j1}^2 L/4E$, $s_{ij} \equiv \sin\theta_{ij}$, $c_{ij} \equiv \cos\theta_{ij}$, and $J \equiv s_{23}c_{23} s_{13} c_{13}^2 s_{12} c_{12}\sin\delta_{\rm CP}$ is the Jarlskog invariant~\cite{Jarlskog:1985ht}. Effects of propagation through matter are given by the matter potential~\cite{Wolfenstein:1977ue},
\begin{equation}
a = \frac{G_F n_e}{\sqrt{2}} \approx \frac{1}{3500\ \mathrm{km}} \left(\frac{\rho}{3.0\ \mathrm{g/cm}^3}\right),
\end{equation}
where $\rho$ is the density along the path of propagation. For current/planned $\nu_e$-appearance oscillation experiments,
\begin{align}
aL=\left\{ \begin{array}{ll}
0 \quad & \text{Vacuum, any $L$ } \\
0.065 \quad & \text{T2K/T2HK~\cite{Abe:2018uyc}} \\
0.22 & \text{NOvA~\cite{Acero:2019ksn}}\\
0.29  & \text{T2HKK~\cite{Abe:2016ero}}\\
0.35 & \text{DUNE~\cite{Abi:2020evt}}
\end{array} 
\right.,
\end{align}
while $\left\lvert\Delta_{31}\right\rvert \approx \pi/2$ so that oscillations due to the atmospheric mass-squared splitting are maximized.

For antineutrinos, $P_{\overline{\mu}\overline{e}} \equiv P(\overline{\nu}_\mu \to \overline{\nu}_e)$ can be determined by taking Eq.~\eqref{eq:PmeMatter} and replacing $\delta_{\rm CP} \to -\delta_{\rm CP}$ as well as $(aL) \to -(aL)$. In Fig.~\ref{fig:BiProbPP} we display how the oscillation probabilities $P_{\mu e}$ and $P_{\overline{\mu}\overline{e}}$ vary at NOvA (left panel) and T2K (right) baselines/energies. We assume fixed $L = 810$ km (left) and $295$ km (right), as well as $E = 2.1$ GeV (left) and $0.6$ GeV (right). We fix $\sin^2\theta_{12} = 0.310$, $\sin^2\theta_{13} = 0.022$, and $\Delta m_{21}^2 = 7.53\times 10^{-5}$ eV$^2$~\cite{SKNu2020,Gando:2013nba,Adey:2018zwh}. Ellipses arise by varying $\delta_{\rm CP}$ for different combinations of $\left( \sin^2\theta_{23},\ \Delta m_{31}^2\right)$, obtained from fits to NOvA (blue ellipses) or T2K (red), assuming the MO is normal (solid) or inverted (dashed). We discuss how these points are obtained in the ``results'' section. Fig.~\ref{fig:BiProbPP} also displays measured oscillation probabilities (with statistical uncertainty) as black (current data~\cite{T2KNu2020, NOvANu2020}) and grey (pre-2020 data~\cite{Abe:2018wpn, Acero:2019ksn}) crosses. Comparing older results to the current ones, we see that the measured oscillation probabilities are trending toward the ``IO'' region of this space, where $P_{\overline{\mu}\overline{e}} > P_{\mu e}$.

We also analyze sums/differences of the neutrino and antineutrino oscillation probabilities, $\Sigma P_{\mu e} \equiv P_{\mu e} + P_{\overline{\mu}\overline{e}}$ and $\Delta P_{\mu e} \equiv P_{\mu e} - P_{\overline{\mu}\overline{e}}$. Near $\left\lvert \Delta_{31} \right\rvert \approx \pi/2$, and assuming $(aL) \ll \Delta_{31}$,
\begin{widetext}
\begin{subequations}
\begin{align}
\Sigma P_{\mu e} &\to 8 s_{13}^2 c_{13}^2 s_{23}^2 - 16s_{12} c_{12} s_{13} c_{13}^2 s_{23} c_{23} \sin\delta_{\rm CP}(aL) \frac{\Delta m_{21}^2}{\left\lvert\Delta m_{31}^2\right\rvert} \mathrm{sign}\left(\Delta m_{31}^2\right),\notag \\
\Sigma P_{\mu e} &\approx 0.17 s_{23}^2 - 0.03(aL)s_{23} c_{23} \sin\delta_{\rm CP} \mathrm{sign}\left(\Delta m_{31}^2\right), \label{eq:SigPnum}\\
\Delta P_{\mu e} &\to \frac{32(aL)}{\pi}s_{13}^2 c_{13}^2 s_{23}^2 \mathrm{sign}\left(\Delta m_{31}^2\right) - 8\pi s_{12} c_{12} s_{13} c_{13}^2 s_{23} c_{23} \sin\delta_{\rm CP} \frac{\Delta m_{21}^2}{\left\lvert \Delta m_{31}^2\right\rvert},\notag \\
\Delta P_{\mu e} &\approx 0.22(aL)s_{23}^2 \mathrm{sign}\left(\Delta m_{31}^2\right) - 0.05 s_{23} c_{23} \sin\delta_{\rm CP},\label{eq:DelPnum}
\end{align}
\end{subequations}
\end{widetext}

\begin{figure}[!h]
\begin{center}
\includegraphics[width=0.825\linewidth]{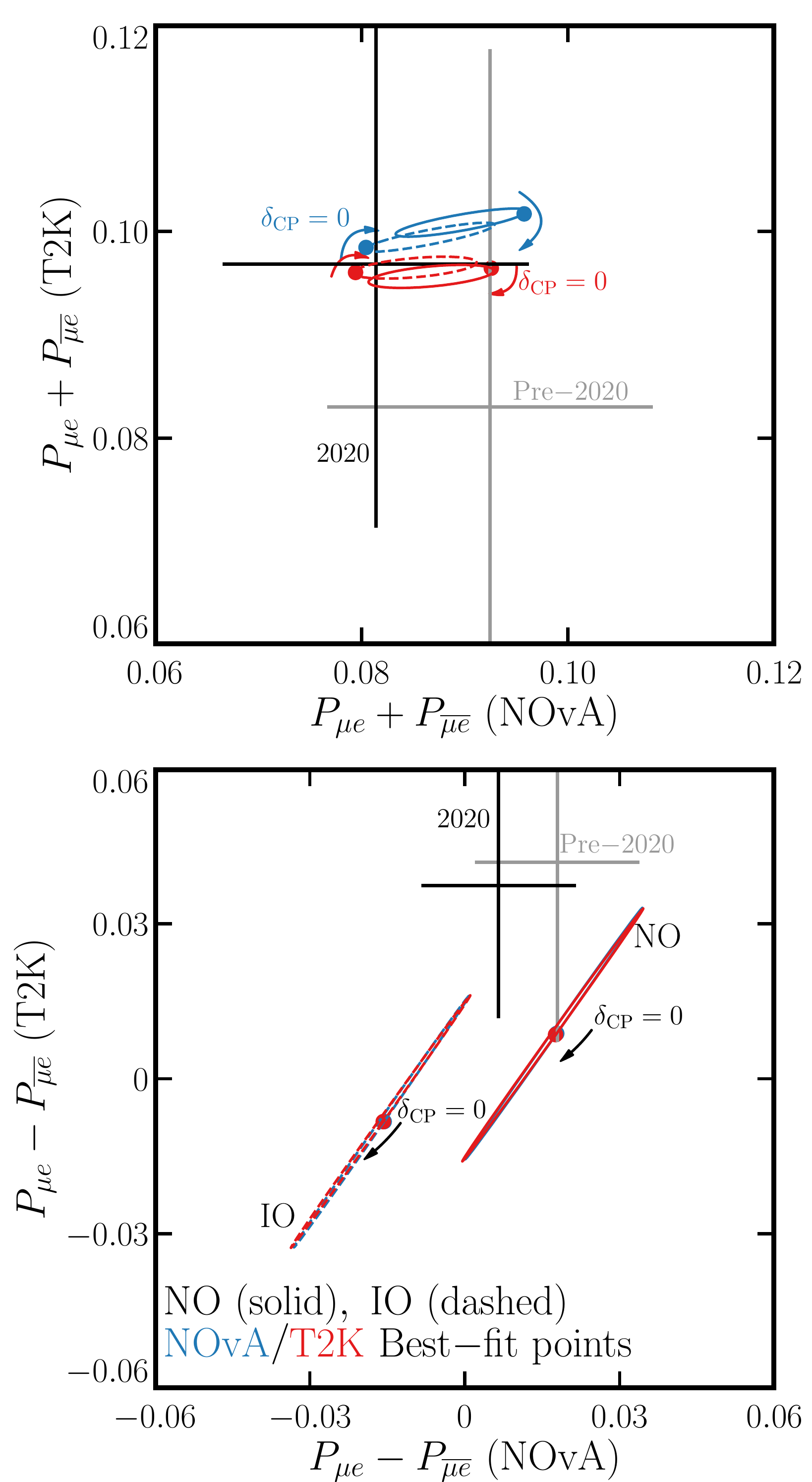}
\caption{Sums (top) and differences (bottom) of oscillation probabilities at NOvA (x-axes) and T2K (y-axes) at fixed $L$ and $E$. Ellipses are generated by varying $\delta_{\rm CP}$ while fixing the other oscillation parameters. Dots denote $\Sigma P_{\mu e}$ (top) and $\Delta P_{\mu e}$ (bottom) for $\delta_{\rm CP} = 0$, with arrows indicating increasing $\delta_{\rm CP}$. Crosses display sums/differences of oscillation probabilities assuming statistically-independent measurements at NOvA/T2K for current (black) and pre-2020 (grey) results.
\label{fig:BiProbSPDP}}
\end{center}
\end{figure}
\noindent where we have used the current best-fit measurements of $\theta_{12}$, $\theta_{13}$, $\Delta m_{21}^2$, and $\left\lvert \Delta m_{31}^2\right\rvert$. Analyzing Eq.~\eqref{eq:SigPnum}, we see that measurements of the sum of oscillation probabilities allow for extracting $s_{23}^2$, while the effects of CP violation and the MO have a small impact. According to Eq.~\eqref{eq:DelPnum}, measurements of $\Delta P_{\mu e}$ can allow for extraction of the MO, octant, and CP violation, however these are all comparable and competing effects. 

We show the sums and differences of oscillation probabilities at NOvA/T2K in Fig.~\ref{fig:BiProbSPDP}, presenting $\Sigma P_{\mu e}$ ($\Delta P_{\mu e}$) in the top (bottom) panel. Extracted measurements of these sums/differences are displayed as black (current) and grey (pre-2020) crosses, assuming statistically-independent measurements of $P_{\mu e}$ and $P_{\overline{\mu}\overline{e}}$ at each experiment. Red/blue ellipses are generated fixing all parameters except $\delta_{\rm CP}$ to the same combinations as in Fig.~\ref{fig:BiProbPP}. In the bottom panel, the impact of the MO is clear -- NOvA requires $\Delta P_{\mu e} > 0$ for NO and $\Delta P_{\mu e} < 0$ for IO. While the separation is not as powerful for T2K (where $(aL)$ is a factor of ${\sim}3$ smaller), the NO prefers larger $\Delta P_{\mu e}$. As with Fig.~\ref{fig:BiProbPP}, we see that current data have moved towards favoring IO for both T2K and NOvA. In what follows, we quantify these effects, with fits to T2K/NOvA to determine their individual and joint preferences for the MO.

\textbf{Analysis.} ---
For T2K, we consider the latest results, equivalent to $1.97(1.63)\times 10^{21}$ protons-on-target in neutrino (antineutrino) mode~\cite{T2KNu2020}. T2K observes a total of 108 (16) $\nu_e$ ($\overline{\nu}_e$) like events, and 318 (137) $\nu_\mu$ ($\overline{\nu}_\mu$) like events. We classify the data in the same categories as T2K, muon-ring ($1R\mu$) and electron-ring ($1Re$) events in neutrino and antineutrino modes, plus $\nu_e - CC1\pi$ events in neutrino mode~\cite{T2Kphdth}.  We define a log-likelihood function comparing the expected/observed events (including normalization uncertainties). We consider uncertainties for $1Re$ events of 4.7\% (5.9\%), while for $1R\mu$ we assume 3.0\% (4.0\%) in neutrino (antineutrino) mode. For $\nu_e - CC1\pi$ we use a 14.3\% uncertainty. 

Our analysis of NOvA is similar to T2K -- we reproduce the expected spectra for $\nu_\mu$ and $\overline{\nu}_\mu$ disappearance channels given a set of input parameters, as well as the number of $\nu_e$ and $\overline{\nu}_e$ appearance events. We determine these spectra following the results of Ref.~\cite{NOvANu2020}, which correspond to $13.5(12.5)\times 10^{20}$ proton-on-target-equivalent in neutrino (antineutrino) mode. These spectra and event rates enter a Poissonian log-likelihood function as with T2K -- NOvA observes a total of 82 (33) $\nu_e$ ($\overline{\nu}_e$) like events, and 211 (105) $\nu_\mu$ ($\overline{\nu}_\mu$) like events in this dataset. Since NOvA is statistics-limited, we do not incorporate any systematic uncertainties at this stage of our analysis. See Appendices~\ref{app:T2K} and~\ref{app:NOvA} for details of our T2K and NOvA analyses, respectively.

In order for the long-baseline experiments to measure $\nu_e$ and $\overline{\nu}_e$ appearance, external information is required. For simplicity, we fix the parameters that T2K/NOvA do not measure to their current best-fit values from other experiments: $\sin^2\theta_{12} = 0.307$, $\sin^2\theta_{13} = 0.0218$, $\Delta m_{21}^2 = 7.53 \times 10^{-5}$ eV$^2$~\cite{SKNu2020,Gando:2013nba,Adey:2018zwh}. Allowing these parameters to be free within their allowed ranges in this analysis does not have a significant impact on the resulting interpretation of the long-baseline results. In our conclusions, we discuss how additional reactor antineutrino measurements can impact the MO determination. Finally, we include the $\Delta \chi^2$ map from Ref.~\cite{Abe:2017aap} which we refer to as ``SK18'' henceforth.

\begin{figure}
\begin{center}
\includegraphics[width=\linewidth]{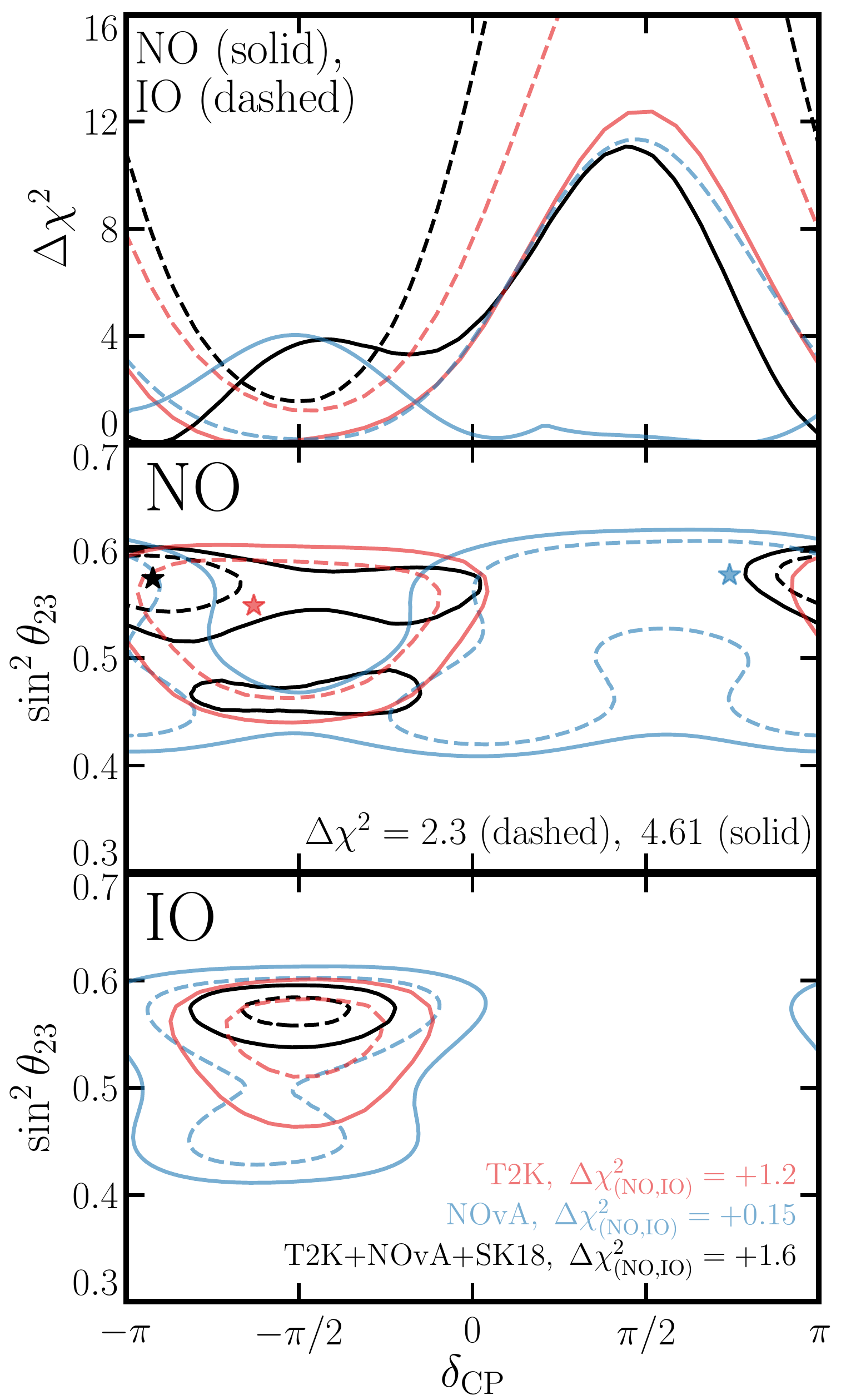}
\caption{Results of our fit of $\delta_{\rm CP}$ and $\sin^2\theta_{23}$. Top: $\Delta \chi^2$ as a function of 
 $\delta_{\rm CP}$  for fixed MO. Middle (Bottom): $\delta_{\rm CP}$ versus $\sin^2\theta_{23}$  for the NO(IO).
Except for the top panel, all contours are $\Delta\chi^2 = 2.3$ (solid) and $4.61$ (dashed)-- black lines indicate a joint fit of T2K/NOvA/SK18, where blue (red) indicate a fit to NOvA (T2K) alone. Corresponding stars indicate the best-fit point of each fit. The text indicates the relative preference for mass ordering by each fit.
\label{fig:DCPS23}}
\end{center}
\end{figure}

\textbf{Results.} --- We perform three different analyses and compare their results. First, we perform a joint analysis of T2K/NOvA/SK18. As we have observed before, cf Fig.~\ref{fig:BiProbPP}, without SK18, this fit results in a mild preference for the IO ($\Delta \chi^2_{\rm (NO,IO)} = -2.6$) and a preference for $\delta_{\rm CP} \approx -\pi/2$, maximal CP violation. When SK18 is included, this preference changes to $\Delta \chi^2_{\rm (NO,IO)} = 1.6$, and  $\delta_{\rm CP} \approx -3$ is favored. Fig.~\ref{fig:DCPS23} presents the results of this fit in black, where the top panel displays the one-dimensional $\Delta \chi^2$ as a function of $\delta_{\rm CP}$ (marginalized over the other five oscillation parameters) when we fix ourselves to be in the NO (solid black line) or IO (dashed black line). The middle (bottom) panel presents two-dimensional measurement contours of constant $\Delta\chi^2$ relative to the best-fit point ($\Delta\chi^2 = 2.3$, dashed, and $4.61$, solid) of $\delta_{\rm CP}$ vs. $\sin^2\theta_{23}$, assuming NO (IO). The best-fit point, $\sin^2\theta_{23} \approx 0.57$, $\delta_{\rm CP} \approx -3$, NO, is shown as a star in the middle panel. We find that the combined results are consistent with the hypothesis that CP is conserved ($\delta_{\rm CP} = 0$ or $\pm\pi$) at $\Delta\chi^2 <1$.

The other two fits we perform are with only T2K or only NOvA. Results of these two fits are shown in Fig.~\ref{fig:DCPS23} for NOvA (blue) and T2K (red). Each fit results in a small preference for NO over IO, again, as discussed cf. Fig.~\ref{fig:BiProbPP}, with T2K giving $\Delta \chi^2_{\rm (NO,IO)} = 1.2$ and NOvA giving $\Delta \chi^2_{\rm (NO,IO)} = 0.15$. The red/blue stars in the middle panel of Fig.~\ref{fig:DCPS23} represent the best-fit points\footnote{In NO (IO), T2K prefers $s_{23}^2 = 0.55$, $\Delta m_{31}^2 = +2.56 \times 10^{-3}$ eV$^2$, $\delta_{\rm CP} = -1.98$ ($0.55, -2.46 \times 10^{-3}$ eV$^2$, $-1.49$). NOvA prefers $s_{23}^2 = 0.58$, $\Delta m_{31}^2 = 2.52 \times 10^{-3}$ eV$^2$, $\delta_{\rm CP} = 2.32$ ($0.57$, $-2.41 \times 10^{-3}$ eV$^2$, $-1.51$) for the NO (IO).} of these fits. The two NO best-fit regions are somewhat in tension, leading to a joint T2K/NOvA fit preferring maximal CP violation, but IO. We show a version of Fig.~\ref{fig:DCPS23} without SK18 in our Appendix~\ref{app:NoSK}.

\begin{table}
\caption{MO preference by an experiment or combination of experiments. The plus (minus) sign indicates preference for NO (IO). We also present the best-fit value of $\delta_{\rm CP}$ and the exclusion of CP conservation.\label{tab:MOPref}}
\begin{tabular}{c|c|c|c}
Experiment(s) & $\Delta\chi^2_{\rm (NO,IO)}$ & Best-fit $\delta_{\rm CP}/\pi$ & $\Delta \chi^2_{\rm CPC}$\\ \hline \hline
T2K & $+1.2$ & $-0.6$ & 2.9\\
NOvA & $+0.15$ & $+0.7$ & 0.33\\
SK18/SK20 & $+3.4$/$+3.2$ & $-0.7/{-}0.6$ & 0.35/0.81\\
T2K + NOvA & $-2.6$ & $-0.5$ & 2.8\\
T2K + SK18 & $+5.7$ & $-0.6$ & 3.0\\
NOvA + SK18 & $+3.6$ & $+0.9$ & 0.53\\ \hline
T2K+NOvA+SK18 & $+1.6$ & $-0.9$ & 0.31
\end{tabular}
\end{table}
Table~\ref{tab:MOPref} summarizes the MO preference by each experiment or combination of experiments we consider, as well as the updated SK 2020 result~\cite{SKNu2020} -- as these results are not yet published, we do not have a $\Delta \chi^2$ map for this to perform a complete joint fit. We comment on a prospective T2K/NOvA/SK20 fit below.

\textbf{Discussion \& Conclusions.} --- The neutrino mass ordering remains one of the largest mysteries in the Standard Model. Previously, experimental data seemed to be preferring the normal mass ordering, $m_1 < m_2 < m_3$, corresponding to the same ordering of the charged fermions of the Standard Model. However, as we have shown, this evidence is waning given updated results from T2K and NOvA, specifically when the two are combined in a joint fit. With SK18, a mild preference for the normal ordering is obtained. However, preliminary updated results from SK could reduce this preference -- Ref.~\cite{SKNu2020} (with updated data) prefers NO at a lower strength than in Ref.~\cite{Abe:2017aap}. As demonstrated in this work, the interplay between $\sin^2\theta_{23}$, $\delta_{\rm CP}$ and the mass ordering is complex, and a complete fit of T2K+NOvA+SK20 is surely warranted.

In this work, we focused on the experiments with recent updates, specifically the long-baseline experiments T2K and NOvA. Reactor antineutrino experiments such as Daya Bay and RENO, when combined with long-baseline disappearance measurements can provide sensitivity to the MO through precise measurements of the effective mass-squared splittings in the two oscillation environments~\cite{Nunokawa:2005nx,Minakata:2006gq}. As shown in Ref.~\cite{Esteban:2020cvm}, inclusion of reactor data pushes the preference further toward NO at the final level of $\Delta \chi^2_{\rm (NO,IO)} \approx 7.1$ ($2.7$ without SK18). While the resulting preference is in the direction of NO, it is very intriguing that the different mechanisms of probing the MO (long-baseline appearance vs. disappearance measurements) yield different MO determinations. Given all of these effects, we must conclude that more data is required before the neutrino mass ordering can be definitively determined.

The importance of this result cannot be understated. 
If neutrinos follow the inverted ordering, this has far-reaching consequences. 
If, in addition, neutrinos are Majorana fermions, there exists some minimum mass relevant for neutrino-less double beta decay. 
If the inverted ordering is true and neutrino-less double beta decay remains unobserved by upgraded experiments,
we can determine that neutrinos are Dirac fermions. 
Moreover, measurements of the cosmic microwave background and the matter power spectrum allow us to infer the sum of the neutrino masses. 
If neutrinos follow the inverted ordering, their sum is at least ${\sim}100$ meV, while for normal ordering $60$ meV. 
The lower limit for the inverted ordering is attainable by next-generation experiments.
Further, experiments that measure neutrino masses via kinematic effects, such as KATRIN~\cite{Aker:2019uuj} and Project 8~\cite{Esfahani:2017dmu}, could also be impacted by the mass ordering; the minimum effective electron neutrino mass is about 50~meV for the inverted ordering as opposed to 9~meV for normal ordering.
The mass ordering may also play an important role in the potential observation of relic neutrinos from the early universe by the proposed PTOLEMY experiment~\cite{Betti:2019ouf}.

It could be that statistical fluctuations in the data, moving in the same direction in the bi-probability planes of Fig.~\ref{fig:BiProbPP}, are the cause for this vanishing preference for normal ordering. 
This highlights the importance of two things. 
First, the accumulation of more data. As T2K and NOvA continue to run, their statistical uncertainties will decrease, and thus will become more robust against statistical fluctuations. 
Second, the need for the future experiments that will definitively pin down the neutrino mass ordering. 
Between JUNO's long-baseline reactor antineutrino measurements, and DUNE and T2HK's long-baseline high-energy oscillation and atmospheric oscillation measurements (which take different approaches to determine the ordering~\cite{Ternes:2019sak}), we will be able to determine the neutrino mass ordering absolutely.


\vspace{0.7cm}
\centerline{\bf Acknowledgments}
\vspace{0.2cm}
KJK, PANM, SJP, and YFPG are supported by Fermi Research Alliance, LLC under contract DE-AC02-07CH11359 with the U.S. Department of Energy. RZF is supported by CNPq and FAPESP.
This project has received support from the European Unions Horizon 2020 research and innovation programme under the Marie Sklodowska-Curie grant agreement No 690575 and No 674896.

\textbf{Note added:} After the completion of this manuscript, it came to our attention that similar considerations on the preference for inverted ordering of recent T2K and NOvA data were made by Jo{\~a}o Coelho at the Neutrino 2020 conference Slack channel \texttt{\#talks\_lbl\_nus\_beams}.


\onecolumngrid

\appendix
\section{T2K simulation details}\label{app:T2K}
 
As stated in the main text, the data sample is classified according to the observed events in the Far Detector: muon-like (1R$\mu$) and electron-like (1R$e$) ring events, together with $\nu_e\  \rm{CC}1\pi-$like (1R$e$+1$\pi$) events in neutrino mode only. The expected number of events in each category $\beta$ and bin $j$, $N_{\beta j}$ are computed as \cite{Esteban:2018azc}
\begin{widetext}
\begin{align}\label{eq:events}
    N_{\beta j} = N_{{\rm bkg}, \beta} + \int_{E_j}^{E_{j+1}}dE_{\rm rec}\, \int dE_\nu\, {\cal R}(E_{\rm rec},E_\nu)\, \varepsilon(E_\nu)\,\phi(\nu_\mu)\,P(\nu_\mu\to\nu_\beta) \,\sigma_\beta(E_\nu),
\end{align}
with $N_{{\rm bkg}, \beta}$ the number of background events in the sample $\beta$, ${\cal R}(E_{\rm rec},E_\nu)$ the energy resolution function (taken as Gaussian), $\varepsilon(E_\nu)$ the detector efficiency, $\phi(\nu_\mu)$ the initial $\nu_\mu$ flux at the detector site, $P(\nu_\mu\to\nu_\beta)$ the oscillation probability and $\sigma_\beta(E_\nu)$ the correspondent cross section for the sample $\beta$. 
Although eq.~\eqref{eq:events} is written for neutrino events, a similar one can be inferred for antineutrinos. We consider the SK detector, located at 295 km and $2.5^{\circ}$ off-axis from the neutrino source, with a fiducial volume of 22.5 kt. The initial fluxes from J-PARC and the cross sections are obtained from Ref.~\cite{T2Kphdth}. 

The test statistics is taken as Poissonian with priors on pull and oscillation parameters,
\begin{align}\label{eq:chisq}
    \chi^2=&\ 2\sum_{\beta}\sum_{j=1}^{b_\beta}\left\{(1+f_\beta)N_{\beta j} - d_{\beta j} + d_{\beta j}\log\left(\frac{d_{\beta j}}{(1+f_\beta)N_{\beta j}}\right)\right\}+\sum_{\beta}\left(\frac{f_\beta}{\sigma_\beta}\right)^2,
\end{align}
\end{widetext}
where $d_{\beta j}$ is the T2K data, taken from~\citep{T2KNu2020}, in the sample $\beta$ and reconstructed energy bin $j$; note that the number of bins $b_\beta$ depends on the sample. The factors $f_\beta$ are pull parameters for the normalization systematic uncertainties of the neutrino data sets, see Tab.\ \ref{tab:ParNormSys}. 

\begin{table*}[ht]
\caption{\label{tab:ParNormSys} Uncertainties and number of bins the different samples considered in out T2K simulation.}
\begin{ruledtabular}
	\begin{tabular}{lcc}
		\toprule\toprule
         Sample & Uncertainty & Number of bins $b_\beta$\\ \midrule
         1R$\mu$ $\nu-$mode & $3.0\%$ & 40\\
         1R$\mu$ $\bar{\nu}-$mode & $4.0\%$ & 40\\
         1R$e$ $\nu-$mode & $4.7\%$ & 25\\
         1R$e$ + 1$\pi$ $\nu-$mode & $14.3\%$ & 25\\
         1R$e$ $\bar{\nu}-$mode & $5.9\%$ & 25\\
		 \bottomrule
	\end{tabular}
	\end{ruledtabular}
\end{table*}

\onecolumngrid
\section{NOvA simulation details}\label{app:NOvA}
Our simulation for NOvA follows the same strategy described above for T2K, where we simulate event spectra as a function of reconstructed neutrino energy (and including backgrounds) according to Eq.~\eqref{eq:events}. While our T2K analysis includes five channels (separating out the single-pion events in the $\nu_e$ appearance mode), our NOvA analysis only includes four -- the two disappearance channels (one in neutrino mode and one in antineutrino mode) and the two appearance channels.

Given a set of oscillation parameters, we therefore can produce expected event spectra for neutrino appearance and disappearance. After calculating the expected event spectrum for (anti)neutrino disappearance, we bin our data to match the binning in Ref.~\cite{NOvANu2020}. These binned data are compared against the data extracted from Ref.~\cite{NOvANu2020} using a Poissonian log-likelihood function like the one given in Eq.~\ref{eq:chisq} to determine the $\chi^2$ given this simulated expected event spectrum. As explained in Ref.~\cite{NOvANu2020}, NOvA's measurements are currently statistics-limited so we do not incorporate any systematic uncertainties in our fit to NOvA data.

For the electron neutrino appearance channels in neutrino and antineutrino modes, we perform a rate-only measurement, comparing the expected number of events in each channel (given some set of input oscillation parameters) to the measured event rates, 82 and 33 in neutrino and antineutrino modes, respectively.

\onecolumngrid
\section{Results of T2K/NOvA Fits without Super-Kamiokande}\label{app:NoSK}
\begin{figure}
\begin{center}
\includegraphics[width=0.5\linewidth]{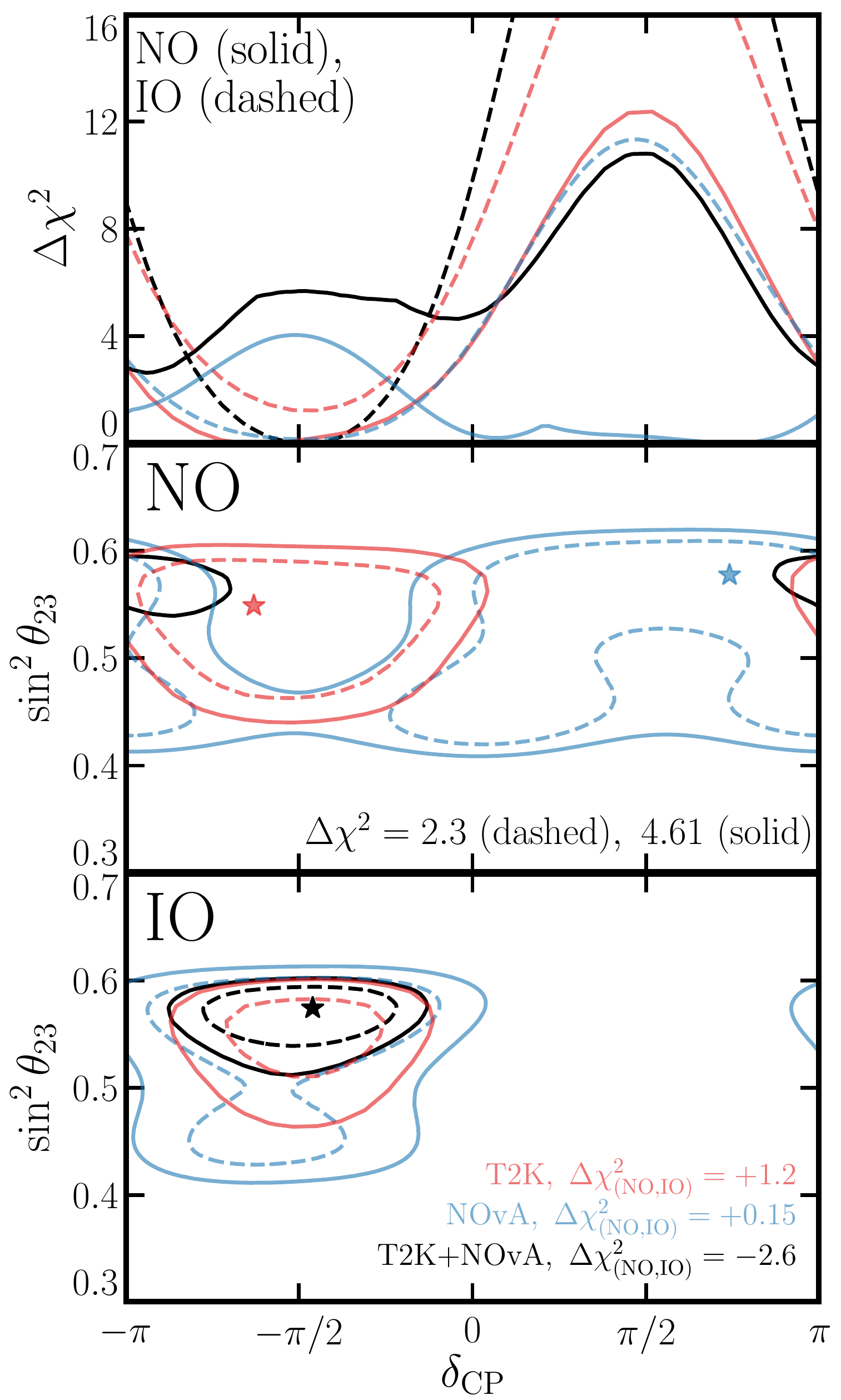}
\caption{Results of our fit of the oscillation parameters $\delta_{\rm CP}$ and $\sin^2\theta_{23}$. In the top panel we show $\Delta \chi^2$ as a function of 
 $\delta_{\rm CP}$  for a fixed MO. In the middle (bottom) panel we show $\delta_{\rm CP}$ versus  $\sin^2\theta_{23}$   for the  NO(IO).
Except for the top panel, all contours are  $\Delta \chi^2 = 2.3$ (dashed) and $4.61$(solid) -- black lines indicate a joint fit of T2K/NOvA, where blue (red) indicate a fit to NOvA (T2K) alone. The corresponding stars indicate the best-fit point of each fit, and the text indicates the relative preference for mass ordering by each fit ($\Delta\chi^2_{\rm(NO,IO)} \equiv \chi^2_{\rm(min,IO)}-\chi^2_{\rm(min,NO)}$).
\label{fig:DCPS23_NoSK}}
\end{center}
\end{figure}

In Fig.~\ref{fig:DCPS23} of our main text we presented the results of fits to NOvA and T2K's recently-updated data, as well as a combined fit to T2K, NOvA, and Super-Kamiokande's published result from Ref.~\cite{Abe:2017aap}, using the $\Delta \chi^2$ map from that publication. In this section, we repeat the exercise of Fig.~\ref{fig:DCPS23} with a joint T2K/NOvA fit without Super-Kamiokande. This is shown in Fig.~\ref{fig:DCPS23_NoSK}.

As discussed in the main text, the combination of T2K and NOvA prefer the inverted mass ordering over the normal at the $\Delta \chi^2_{\rm (NO,IO)} = -2.6$ level\footnote{Since we are displaying contours of $\Delta \chi^2 = 2.6$ and $4.61$ in the center and bottom panels, no dashed black contour appears in the middle panel where we are fixing the mass ordering to be normal. This is because the combination of T2K and NOvA in our analysis disfavors the NO at stronger than $\Delta \chi^2 = 2.3$.}. Their combination prefers maximal CP-violation, $\delta_{\rm CP} \approx -\pi/2$, and the upper octant $s_{23}^2 > 1/2$. The marginalized one-dimensional $\Delta \chi^2$ lines in the top panel of Fig.~\ref{fig:DCPS23_NoSK} allow us to determine T2K on its own (with $\sin^2\theta_{13}$ fixed) can exclude a small interval of $\delta_{\rm CP} \approx \pi/2$ at $>3\sigma$ CL. However, once NOvA is included, the interval shrinks (note that near $\delta_{\rm CP} \approx \pi/2$, the exclusion of the red solid line is higher than that of the black solid line). According to our results, the combination of T2K and NOvA can only exclude the hypothesis that CP is conserved ($\delta = 0$ or $\delta = \pm\pi$) at roughly the $\Delta\chi^2 \approx 3$ level.

\onecolumngrid

\bibliographystyle{apsrev4-1}
\bibliography{refs}

\end{document}